\documentstyle[epsf,12pt]{article}
\newcommand{\be}{\begin{equation}}
\newcommand{\ba}{\begin{eqnarray}}
\newcommand{\ee}{\end{equation}}
\newcommand{\ea}{\end{eqnarray}}

\newcommand{\eV}{\;{\mathrm eV}}

\newcommand{\GeV}{\;{\mathrm GeV}}

\newcommand{\cm}{\;{\mathrm cm}}
\newcommand{\sterad}{\;{\mathrm sterad}}
\newcommand{\simlt}{\stackrel{<}{{}_\sim}}

\newcommand{\ol}{\overline}

\newcounter{currequation}
\begin{document}
\title{The Decay of Dark-Matter Inflatons Can Produce Very Energetic 
Cosmic Rays}
\author{Saul Barshay and Georg Kreyerhoff\\
III.~Physikalisches Institut A\\
RWTH Aachen\\
D-52056 Aachen, Germany}
\maketitle
%\vspace{-9cm}
%\flushright {PITHA 98/17}
%\vspace{8cm}\par
\abstract{We have shown that inflatons with a mass which is calculated
to be of the order of $10^{10}\GeV$ can constitute a dominant part
of dark matter. They can decay uniquely into a neutrino $(\nu_\tau$) and antineutrino
with a lifetime calculated to be greater than the present age of the universe.
We show here that these neutrinos can give rise to a localized bump in the
primary energy spectrum of extensive air showers. The structure would 
appear above the sharply-falling spectrum at the expected GZK cut-off
energy for primary protons $\sim 5\times 10^{19}\eV$. The 
flux which is necessary to compensate for the small cross section is,
in principle, attainable, with distant structures also as possible significant
sources, and such events might be observable in coming experiments.}
\section{Introduction}
It is possible that dark matter is predominantly in the form of inflatons
\cite{ref1}. These have a mass which is calculated \cite{ref1}
to be of the order
of $10^{10}\GeV$, and have a lifetime which is calculated \cite{ref1}
to be of the
order of $10^3t_0$ or greater, where $t_0\cong 4\times 10^{17}\;\sec$ is the
present age of the universe. We have calculated explicit potentials
which exhibit features characteristic of an inflationary period in
the early universe\cite{ref2}. These potentials have a calculated
maximum and calculated minimum\cite{ref1,ref2}. This structure is
the result of radiative corrections to the tree-level potential.
The radiative corrections are calculated using the renormalization-
group equations in a (chiral-like) dynamical model involving only
the scalar inflaton field, a pseudoscalar field, and a neutral
lepton\cite{ref1,ref2}. The inflaton decays uniquely\cite{ref1} into a
neutrino $(\nu_\tau)$ and an antineutrino, with energies of the order of $10^{10}\GeV$. 
In this paper, we show that these neutrinos
can give rise to a localized bump in the primary energy spectrum
of extensive air showers. The structure would appear above the
sharply-falling spectrum at the expected Greisen-Zatsepin-Kuzmin
cut-off energy \cite{ref3,ref4,ref5} for primary nucleons, $\sim 5\times
10^{19}\eV$. The essential idea was put forward in \cite{ref1}. Here
we give numerical examples of a localized ``bump'' structure. We estimate
the matter of whether a sufficiently large flux of tau neutrinos is,
in principle, attainable, in order to compensate for the small neutrino-air
interaction probability. We remark upon zenith-angle dependence of air-shower events.
\section{Representation of the data}
In Fig.~1, new data from the AGASA experiment is reproduced \cite{ref7a}.
For a qualitative comparison, we have calculated the curves from
the following phenomenological representation, which involves
adding a local structure to the falling spectrum near to the GZK
cut-off energy $\sim 5\times 10^{19}\eV$.
\be
E^3I\cong\left\{ (10^{24.6})\left(\frac{10^{18}}{E}\right)^{0.2} +
 (E^3 10^{-14.6})\frac{e^{-\frac{(E-E_0)^2}{2\sigma^2}}}{\sqrt{2\pi}\sigma}
 \right\}
\ee
for $10^{18}\le E\le 10^{20.5}$. In eq.~(1), E is the primary energy in eV
and I represents a flux in $({\mathrm m}^2-\sec-{\mathrm sterad}-\eV)^{-1}$,
\underline{times} a (total) interaction probability for a primary neutrino 
traversing the
(full) atmosphere (i.~e.~the fraction of the flux which interacts to
produce an extensive air shower). The first term is an \underline{approximate}
representation of the data in Fig.~1 in the interval $10^{18}\le E\le 10^{19}$.
For this part of the spectrum presumably due to protons with interaction
probability in the atmosphere of about unity, the flux (times $E$)
at $E\cong 10^{18}$ is normalized to about $4\times 10^{-12}\;
 ({\mathrm m}^2-\sec-{\mathrm sterad})^{-1}$. The second term represents
the hypothetical localized structure. The neutrino energy is $E_0$ and
to illustrate the possible effect, we have assumed a Gaussian distribution
of apparent values of $E$, characterized by a spread $\pm\sigma$ about
$E_0$. For this part of the spectrum, the flux (times $E$) times
interaction probability is normalized to about 
$1.4\times 10^{-15} ({\mathrm m}^2-\sec-{\mathrm sterad})^{-1}$ at 
$E=E_0\cong \sqrt{2}\sigma=10^{19.5}$.
The dashed curve is for these values of $ E_0,\sigma$. The solid curve is
for $E_0\cong\sigma=10^{19.5}$. The dashed-dotted curve
is for $E_0=8\times 10^{19}, \sigma=5\times 10^{19}$. At the upper edge of
the structure we have removed the (then comparable) contribution of the first term in 
eq.~(1), so as to roughly simulate the GZK cut-off. The arrow indicates the point to which the
structure then falls. An expected slight excess just before the GZK cut-off is {\underline{not}}
included in our parametrization in eq.~(1). In Fig.~2, the contribution from
the first term in eq.~(1) is (abruptly) removed at $E=10^{19.7}$, so as to simulate
the GZK cut-off at this point. We emphasize that Figs.~1,2 are to be considered
as {\underline{examples}} of the hypothetical ``bump'' structure; they are not a
representation of the present limited data (which can be even be represented
by a flat line, above $10^{19}\;\eV$).
\section{The flux of neutrinos from the decay of dark-matter inflatons}
The qualitative comparision to present data provided by the curves in Figs.~1,2
suggests a flux times interaction probability somewhat less than
of $10^{-15} ({\mathrm m}^2-\sec-{\mathrm sterad})^{-1}$.
Consider a very energetic neutrino, $E \sim 10^{10}\GeV$. The
neutrino-air cross section can be greater than $10^{-32}\;
{\mathrm cm}^2$. The mean-free path $l_0$, is then of the order of $\frac{1}{2}\times 
10^{13}{\mathrm cm}$. The fraction of neutrinos which interact in a
characteristic atmospheric length $l_{\mathrm atm}$ of roughly 25 km
is (conservatively) about $\left(\frac{l_{\mathrm atm}}{2l_0}\right)\cong 2.5\times 10^{-7}$. 
We write the flux as approximately ( for $\tau \gg t_0$, and $E_{\nu_\tau}\simlt \frac{m}{2}$)
\be
I_{\nu_\tau} \cong \frac{1}{4\pi}\frac{\rho_{\mathrm CDM}}{m}\frac{L}{\tau}
\ee
In eq.~(2), $\rho_{\mathrm CDM}$ is the energy density of cold dark matter
from inflatons of mass m, which have a lifetime $\tau$.
We have \underline{calculated}\cite{ref1} the following estimates: $\rho_{\mathrm CDM}
\cong 2\times 10^{-47}\GeV^4$, $m\cong (6\times 10^9-5\times 10^{10})\GeV$,
$\tau \sim 10^{21}-10^{23}\;\sec$. Using $L\sim 10^{28}\cm$ for the inclusion of
distant structures as possible diffuse sources \cite{ref6a}, and $m\cong 5\times 10^{19}\eV$
(for which $\tau\sim 10^{23}\sec$ \cite{ref1}), eq.~(2) gives
\be
I_{\nu_\tau} \sim 0.4\times 10^{-8}  ({\mathrm m}^2-\sec-{\mathrm sterad})^{-1}
\ee
This suggests that a sizeable flux is, in principle, attainable (also
with smaller effective $L$ compensated by larger effective 
$\rho_{\mathrm CDM})$. There is a small dimunition of neutrino energy here due to red-shift $z<1$.
Diminished inflaton decay at very early times tends to reduce contributions (a lower energies here)
from very large $z$.\par
It is instructive to compare our results with those in a recent
paper\cite{ref7}\footnote{Another recent paper\cite{ref8} has considered the possible connection
between the strongly-interacting decay products of hypothetical, very long-lived
cold dark matter and the highest energy cosmic-ray events. In general, the primaries
are affected by the GZK cut-off, for distant sources.}
which invokes  strongly-interacting
energetic decay products of a hypothetical tiny component of very long-lived cold dark
matter as the origin of the highest energy air showers. First, the
statement\cite{ref7} is made that the highest energy primaries are predominantly
$\gamma$-rays. Second, a broad spectrum of primary energies extends to 
far above $10^{20}\eV$, where
there are no events (at present). Third, the required flux for $E>10^{19}\eV$
is obtained from a formula like eq.~(2) with the following changes:
\begin{itemize}
\item[(a)] $\rho_{\mathrm CDM}$ is taken to be $\cong 0.3\GeV/{\mathrm cm}^3$
(a ``dark-matter halo'' density in our galaxy), times an a priori very small number 
$\epsilon_X$\cite{ref7}. Without the factor $\epsilon_X$, this is $\sim 10^5$
larger than our $\rho_{\mathrm CDM}$.
\item[(b)] This is partly compensated by the ad hoc choice of $m=(10^{13}-10^{14})\GeV$.
\item[(c)] $L$ is taken as a halo length scale of $3\times 10^{23}$ cm. For $\tau\sim
3\times 10^{21}\sec$, these changed factors compensate to 
yield $\sim \frac{\epsilon_X}{(l_{\mathrm atm}/2l_0)}$ relative to
our result for $I_\nu \times ({\mathrm interaction\;\; probability})$. The ad hoc
parameter \cite{ref7} $\epsilon_X< 2.5\times 10^{-7}$ is thus  essentially taking the
role of the known small probability for neutrino interaction
\footnote{A recent article\cite{ref9} concerning, in particular,
the few cosmic-ray events which appear to have primary energies greater than
$10^{20}\eV$. has proposed that these are due to neutrinos which acquire strong
interactions at c.~m.~collision energies above about $10^{14}\eV$. The argument given\cite{ref9} for
assuming a neutrino-air cross section of a few hundred millibarns involves
the hypothesis of strong, long-range interaction of a neutrino
\underline{coherently} with  the quarks in a nucleon. However,
this type of coherent interaction results in diffraction dissociation\cite{ref10};
the cross section for diffraction dissociation must tend to zero as the target continually
blackens at very high c.~m.~collision energies\cite{ref10,ref11}. This cross section is empirically
a near-constant, small part (some millibarns) of the increasing total $p\ol{p}$
cross section at Tevatron energies and above\cite{ref11}. See also \cite{ref12}.}. (Clearly, we can
have the galactic halo contributing in eqs.~(2,3), with a larger
effective $\rho_{\mathrm CDM}$ at a smaller $L$. One can even imagine MACHOs.)
\end{itemize}
\section{Conclusion}
It is possible that neutrinos from decay of cold dark matter can give
rise to an observable localized ``bump'' structure in the primary energy spectrum
of extensive air showers. This would appear above the sharply-falling
spectrum at the expected (for distant sources) GZK cut-off
energy for primary protons, $\sim 5\times 10^{19}\eV$. With finite
energy resolution, this structure would ensure that there is a smooth
transition near to the GZK cut-off energy to energies just above.\par
Given the stated presence \cite{ref15} of attenuation with zenith angle in
the handful of the highest-energy events, it seems unlikely that these could be all
initiated by (weakly-interacting) tau neutrinos. It is essential to take into
account that there are some protons near to $10^{20}\eV$ (as a tail of the GZK cut-off, at least).
Only for the (presently relatively few) events in the highest-energy bins
would one expect a diminished presence of zenith-angle dependence, if (weakly interacting$^{\mathrm F2}$)
$\nu_\tau$ are initiating events. If the mass of the $\nu_\tau$ is only
$\sim 0.05\eV$ (instead of $\sim 1.8\eV$ as used in \cite{ref1}), then the lifetime
is lengthened by $\sim 10^3$, and the flux in eq.~(3) is reduced to $\sim 4\times 10^{-16}
(\cm^2-\sec-\sterad)^{-1}$.\footnote{For scalar inflatons decaying into nearly
massless neutrinos the lifetime can be longer if the interaction contains $(1+b\gamma_5)$ on
the neutrino wave functions, with $b\to 1$. \cite{ref1}.}
A sizeable flux of $\nu_\tau$ should be observable at the larger detection systems
for very high-energy air showers which are being constructed, and in the detectors
which are constructed specifically for neutrinos (with due account taken of
the hadronic shower that could be an observable aspect of the decay of produced,
lower-energy $\tau$).\par
It is possible that compact, very massive quasi-stellar and galactic-core
entities involve inflatons of mass $\sim 10^{10}\GeV$, i.~e.~of the order
of $10^{10}$ times the nucleon mass. The decay produces very large amounts of
energy in radiation associated with such entities. This could provide a
new approach to the development of these energetic objects with very strong 
gravitational fields. (It is noteworthy that this decay provides a 
nonequilibrium
microscopic process for the direction of time on the scale of the universe).
\section*{Acknowledgement}
S.~B.~thanks Prof.~M.~Nagano from Tokyo for information concerning aspects of
the AGASA data.

\newpage
\section*{Figures}
\begin{figure}[h]
\begin{center}
\mbox{\epsfysize 11cm \epsffile{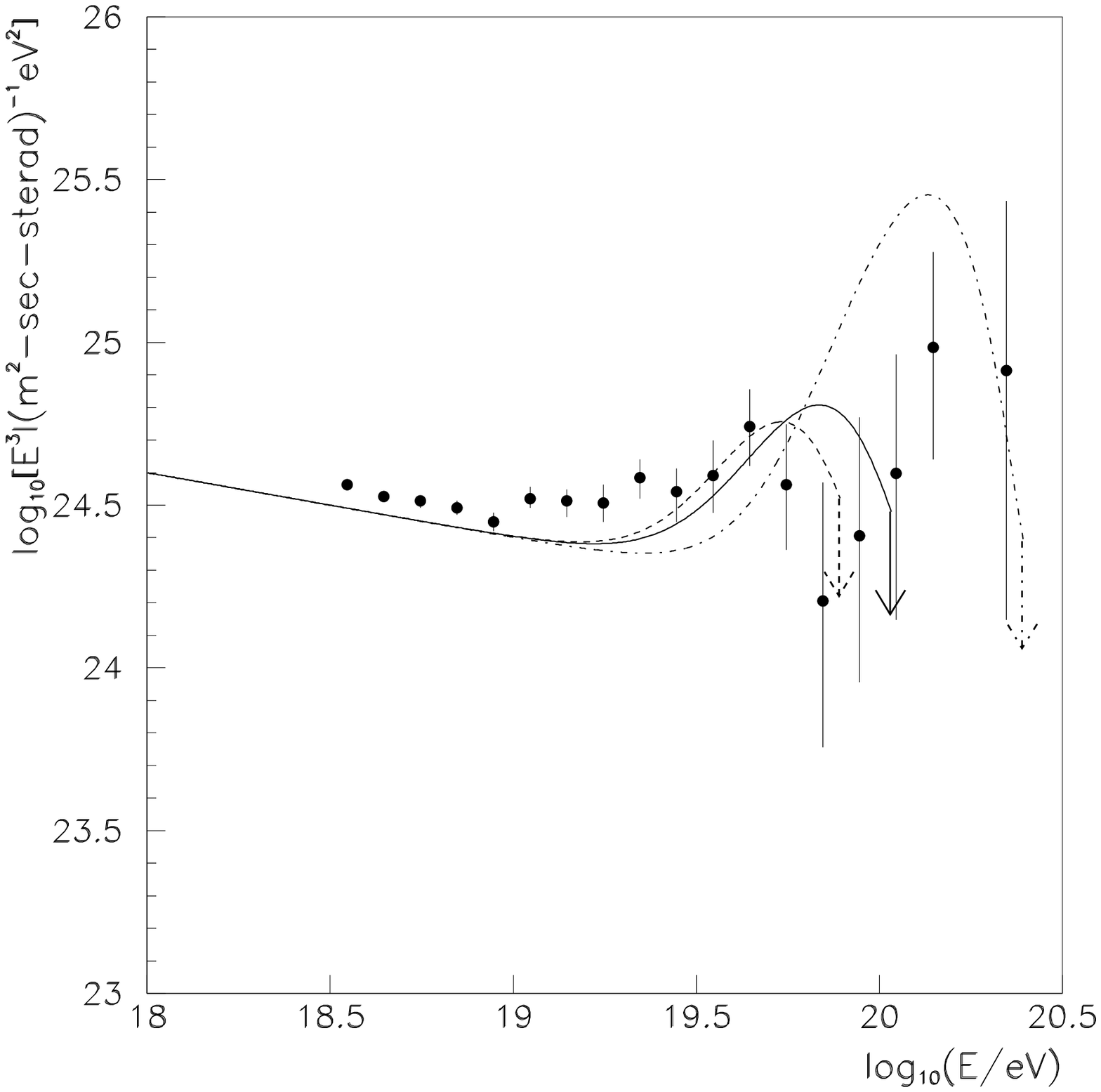}}
\caption{The new data from the AGASA experiment\cite{ref7a} in the interval $10^{18}\eV \le
E\le 10^{20.5}\eV$. The curves are from eq.~(1), with $E_0$ and $\sigma$ as stated in the text
at the end of section 2. (For illustration, we have used a generous spread $\sigma$ in
order to simulate a number of possible uncertainties\cite{ref6}. Possible rapid motions
of decaying inflatons would also contribute to broadening.)}
\end{center}
\end{figure}
\begin{figure}[t]
\begin{center}
\mbox{\epsfysize 12cm \epsffile{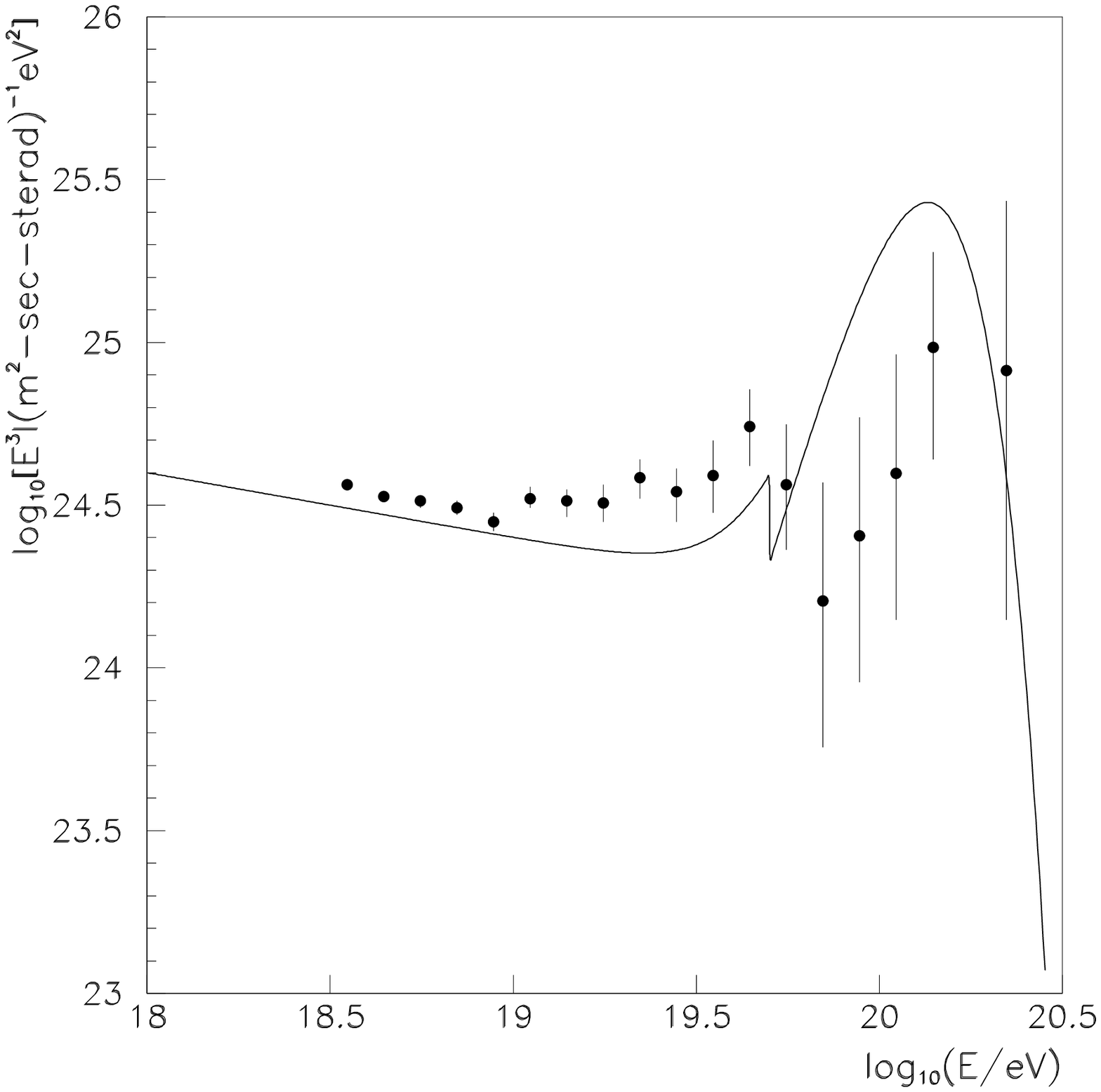}}
\caption{The dashed-dotted curve from Fig.~1, but with the first term in eq.~(1) removed
at $E\protect\cong 10^{19.7}\eV$, so as to roughly simulate the GZK cut-off at this point.
An expected slight ``pile-up'' of events just before the GZK cut-off is not included in our
simple parametrization in eq.~(1), because our concern here is with the idea of a possible
``bump'' structure for the highest-energy events.}
\end{center}
\end{figure}

\begin{thebibliography}{99}
\bibitem{ref1} S.~Barshay and G.~Kreyerhoff, European Physical Journal {\bf C5} (1998) 369
\bibitem{ref2} S.~Barshay and G.~Kreyerhoff, Z.~Phys.~{\bf C75} (1997) 167;
Erratum Z.~Phys.~{\bf C76} (1997) 577
\bibitem{ref3} K.~Greisen, Phys.~Rev.~Lett.~{\bf 16} (1966) 748.
\bibitem{ref4} G.~T.~Zatsepin and V.~A.~Kuzmin, JETP Lett.~{\bf 4} (1966) 78
\bibitem{ref5} N.~Hayashida et.~al., Proc.~of Int.~Symposium {\it Extremely High Energy
Cosmic Rays} (ed.~M.~Nagano) Univ.~of Tokyo (1996) 17\\
V.~Berezinsky, hep-ph/9802351, Feb.~1998
\bibitem{ref6} S.~Yoshida et.~al., Astroparticle Phys.~{\bf 3} (1995) 105\\
S.~Yoshida and Hongyue Dai, astro-ph/9802294, Feb.~1998
\bibitem{ref6a} J.~W. Elbert and P.~Sommers, Ap.~J.~{\bf 441} (1995) 151\\
N.~Hayashida et.~al.,~Phys.~Rev.~Lett.~{\bf 77} (1996) 1000
\bibitem{ref7a} M.~Takeda et.~al.~, astro-ph/9807193, (July 1998)
\bibitem{ref7} V.~Berezinsky, M.~Kachelriess and A.~Vilenkin, 
Phys.~Rev.~Lett.~{\bf 79} (1997) 4302
\bibitem{ref8} V.~A.~Kuzmin and V.~A.~Rubakov, astro-ph/9709187, Sept.~1997
\bibitem{ref9} J.~Bordes, Chan H.-M., J.~Faradini, J.~Pfaudler, and
Tsou S.~T., technical report RAL-TR-97-067, Dec.~1997
\bibitem{ref10} M.~L.~Good and W.~D.~Walker, Phys.~Rev.~{\bf 120} (1960) 1857
\bibitem{ref11} S.~Barshay, P.~Heiliger, and D.~Rein, Z.~Phys.~{\bf C56} (1992) 77  
\bibitem{ref12} V.~S.~Berezinsky and G.~T.~Zatsepin, Phys.~Lett.~{\bf 28B} (1969) 423
\bibitem{ref15} Private communication from Prof.~Nagano, Feb.~1999
\end{thebibliography}
\end{document}